        Kordian Andrzej Smolinski
\documentstyle[12pt,twoside,A4]{article}
\advance\topmargin by .5in

\begin{document}
\thispagestyle{plain}
\title{\marginpar{\vspace{-1in}\hspace{-1in}\small KFT U\L\ 6/93}%
A Model of Quantum Space-Time Symmetry
in Two Dimensions\thanks{Supported by KBN Grant No.~2 0218 91 01.}
\thanks{Talk given during the ``Conference on Quantum Topology'', Manhattan,
Kansas, 24--28 March 1993.}}
\author{Kordian Andrzej Smoli\'nski\thanks{e-mail:
{\tt xmolin@plunlo51.bitnet}}\\
{\em Department of Mathematical Physics, University of
{\L}\'od\'z}\\
{\em ul.~Pomorska 149/153, 90--236 {\L}\'od\'z, Poland}}

\maketitle

\begin{abstract}
An example of a toy model of $D=2$ Minkowski space and Poincar\'e group
with real deformation parameter $q$ is considered. A notion of free
motion is defined. The kinematics and phase-space are constructed and the
``uncertainity'' ralations are found.
\end{abstract}

\section{Introduction}
In the last years many of mathematical physicists try to use the
non-com\-mu\-ta\-tive (``quantum'') geometry \cite{connes}  and quantum group
theory \cite{compact,drinf} in physical problems.  The main reason for this
special interest is a fact that quantum deformation admit more freedom---we
have one or few number-valued parameters in the theory, what allows the better
coincidence with an experimental data.  On the other hand the deformation
parameter $q$ allows to recover the conventional theory in the limit $q\to1$,
what suggest that quantum deformation can be useful as a regularization
procedure.  Therefore the main field for the applications of the quantum group
theory in physical problems are the gauge theories \cite{gauge} and gravity
\cite{majid} as well as the quantum deformations of classical \cite{chlopcy}
and quantum mechanics \cite{ubriaco,qdyn}.

There are also attemptions to quantization of the space-time structures
\cite{luk,carow,castell,kin}.  The basis of this concept is the replacement of
the standard Minkowski space-time by a quantum space and consequently the
Lorentz and Poincar\'e groups by quantum groups.  This is also the main idea of
this paper, which is a study of such deformation with real parameter $q$.

The paper is organized as follows.  In Section 2 we consider the using of the
quantum plane as a quantum space-time.  Section 3 descriebes the groups of
rigid motion of this quantum space-time i.e.\ quantum Lorentz and Poincar\'e
groups.  In section 4 we finally formulate the algebraic structure of the
quantum space time and we find its differential calculus.  In section 5 we
formulate and solve the free motion problem, while in the section 6 we
introduce and investigate kinematic notions, i.e.\ the momenta and the
inertial mass.

\section{Quantum Plane as a Quantum Space-Time}
We start our work with the definition of quantum Minkowski space-time. Let
${\bf C}[x^0,x^1]$ be a free unital algebra over ${\bf C}$ generated by two
elements $x^0$ and $x^1$. In this algebra we choose a two-sided ideal ${\bf
I}[x^0x^1-qx^1x^0]$ generated by the element $x^0x^1-qx^1x^0$ ($q\in{\bf R}$).
The quantum Minkowski space-time is the quotient algebra (Manin's plane
\cite{manin})
\begin{equation}
{\bf E}^{1,1}_q={\bf C}[x^0,x^1]/{\bf I}[x^0x^1-qx^1x^0],       \label{plane}
\end{equation}
equipped with two kinds of ``conjugation'': $\;\bar{}$-operation, which is
anti-linear and $^{\rm T}$-operation, which is anti-involutive. The classical
counterparts of the above two operations are complex conjugation and
transposition respectively. We can construct $^*$-operation (``hermitean
conjugation'') via the composition of the above operations:
$^*=\bar{}\,\circ\,^{\rm T}$, like in the classical case.

Eq.~(\ref{plane}) means that we can identify the elements $x^0x^1$ and
$qx^1x^0$. In the other words, we can write
\begin{equation}
x^0x^1=qx^1x^0,                                               \label{reorder}
\end{equation}
and treate the Eq.~(\ref{reorder}) as the commutation relation of the
generators of ${\bf E}^{1,1}_q$.

An assumption of the hermicity of $x^0$ and $x^1$ leads to the condition
$|q|=1$, it means for real $q=\pm1$, and the requirement of an existence of a
classical limit gives us $q=1$, so we have no deformation. The only one
way to deform the Minkowski space-time with real $q$ is to drop out the
hermicity of space-time generators. But now it arises the question: What is an
observable? The answer is the following. We require that the observable
$\Omega$ should be hermitean, $\Omega^*=\Omega$, because in this case its
spectrum is real and its eigenvalues have a physical meaning. The simplest way
to construct the hermitean elements from $x^\mu$ ($\mu=0,1$) is to take
\begin{equation}
{x^\mu}^2={x^\mu}^*x^\mu.                                   \label{hermicity}
\end{equation}
and treate them as observables describing a time and a position.

It is easy to see that the observable defined by Eq.~(\ref{hermicity}) is in
the classical limit simply the square of $x^\mu$. So it should be positively
defined. It is, if the generators $x^\mu$ are ``real'', i.e.\
\begin{equation}
\bar{x}^\mu=x^\mu.                                            \label{reality}
\end{equation}
Eq.~(\ref{reality}) turns out the Eq.~(\ref{hermicity}) into
\begin{equation}
{x^\mu}^2={x^\mu}^{\rm T}x^\mu.                         \label{transposition}
\end{equation}
In the classical limit, under the irreducibility condition, $x^\mu$ are
represented simply by real numbers.

One could think that the fact that we can measure only squares of time and
position leads to some difficulties in physical meaning of time and position,
because taking a square root of ${x^\mu}^2$ we can obtain only the positive
values. But, in fact, it is nothing strange, since we measure time by counting
{\em periods\/} and spatial coordinate by comparing {\em distances\/}, which
{\em are\/} positive numbers.  Because such situation is in the classical case,
then we can leave the above meaning of the observables describing the spatial
and time coordiantes on the quantum level.

To finish the above considerations we would note that, in fact, our full
Min\-kow\-ski space-time ${\bf E}^{1,1}_q$ is not generated by two elements
$x^0$ and $x^1$, as was mentioned in Eq.~(\ref{plane}), but rather by four
elements, i.e.\ the above generators and their transpositions. However, we will
leave the name Minkowski plane and the notation ${\bf E}^{1,1}_q$ for this, in
fact four-domensional (in geometrical, not algebraical, sense) object.

\section{Quantum Group of Rigid Motions}
Now, we want to look for the group of rigid motions of ${\bf E}^{1,1}_q$ and
call it the quantum $D=2$ Poincar\'e group ($P_q(2)$), while its homogenous
subgroup we will call the quantum $D=2$ Lorentz group ($L_q(2)$).

\subsection{Quantum Lorentz Group}
Let us find firstly the quantum group $L_q(2)$ acting on ${\bf E}^{1,1}_q$ in
co-module
\begin{equation}
\delta:{\bf E}^{1,1}_q\to L_q(2)\otimes{\bf E}^{1,1}_q,        \label{co-mod}
\end{equation}
leaving the commutation relation Eq.~(\ref{reorder}), conjugation structure
Eq.~(\ref{reality}) and the metrics
\begin{equation}
{\mit\Sigma}^2=(x^0)^2-(x^1)^2                                \label{metrics}
\end{equation}
unchanged.

We will look for $L_q(2)$ in the form of matrix group,
\begin{equation}
{\mit\Lambda}=\pmatrix{a&b\cr c&d},
\end{equation}
where the matrix entries belongs to the free involutive algebra generated by
four elements ${\bf A}[a,b,c,d]$ divided by a two-sided ideal ${\bf J}$, which
corresponds to the commutation relation between matrix entries. We assume an
existence of the analogons of $\;\bar{}$- and $^{\rm T}$-operations, and they
are described with using the same notation.

We will find the $L_q(2)$ in the form of $SU_q(1,1)$, the subgroup of
$GL_{p,q}(2)$, described in \cite{schirr}, i.e.\ $c=q^{-1}b^*$ and $d=a^*$. The
condition Eq.~(\ref{reality}) leads to
\begin{equation}
\bar{a}=a,\quad\bar{b}=b.
\end{equation}
Then, we can write explicitly Eq.~(\ref{co-mod})
\begin{equation}
\delta\pmatrix{x^0\cr x^1}
=\pmatrix{a&b\cr q^{-1}b^{\rm T}&a^{\rm T}}
\otimes\pmatrix{x^0\cr x^1},                           \label{lorentz-action}
\end{equation}
and we have the following commutation relations
\begin{eqnarray}
ab&=&qba,                                                    \label{lorentz1}\\
ab^{\rm T}&=&qb^{\rm T}a,\\
bb^{\rm T}&=&b^{\rm T}b,\\
aa^{\rm T}&=&a^{\rm T}a+(1-q^{-2})b^{\rm T}b.                \label{lorentz2}
\end{eqnarray}
We can define the quantum determinant
\begin{equation}
\sigma={\det}_q{\mit\Lambda}=aa^{\rm T}-b^{\rm T}b.               \label{det}
\end{equation}
Note, that $\sigma$ belongs to the center of $L_q(2)$, so we can identify
$\sigma={\bf 1}$ (${\bf 1}$ is the unit of $L_q(2)$).

\subsection{Quantum Poincar\'e Group}
Then we will look for inhomogenous extension of $L_q(2)$. We add to the algebra
${\bf A}[a,b,a^{\rm T},b^{\rm T}]$ a pair of generators $u^0$ and $u^1$ with
their transpositions, describing the translations in directions $x^0$ and $x^1$
respectively. Then we must also extend the ideal ${\bf J}$ by a number of
generators corresponds tho the set of commutation realtions between generators
of $L_q(2)$ and translations and between translations each other.

We assume the action of $P_q(2)$ on ${\bf E}^{1,1}_q$ is the following
\begin{equation}
\delta\pmatrix{x^0\cr x^1\cr {\bf 1}}
=\pmatrix{a&b&u^0\cr q^{-1}b^{\rm T}&a^{\rm T}&u^1\cr0&0&{\bf 1}}
\otimes\pmatrix{x^0\cr x^1\cr {\bf 1}},                        \label{action}
\end{equation}
with
\begin{equation}
\bar{u}^0=u^0,\quad\bar{u}^1=u^1,
\end{equation}
and we find the following commutation relations \cite{inhom}
\begin{eqnarray}
au^0&=&q^{-2}u^0a,                                          \label{poincare1}\\
au^1&=&q^{-1}u^1a,\\
a^{\rm T}u^0&=&q^{-1}u^0a^{\rm T}-(1-q^{-2})u^1b,\\
a^{\rm T}u^1&=&q^{-2}u^1a^{\rm T},
\end{eqnarray}
\begin{eqnarray}
bu^0&=&q^{-2}u^0b,\\
bu^1&=&q^{-1}u^1b,\\
b^{\rm T}u^0&=&q^{-1}u^0b^{\rm T}-q(1-q^{-2})u^1a^{\rm T},\\
b^{\rm T}u^1&=&q^{-2}u^1b,
\end{eqnarray}
\begin{eqnarray}
u^0u^1&=&qu^1u^0,\\
u^0{u^0}^{\rm T}&=&q{u^0}^{\rm T}u^0+q(q^2-1){u^1}^{\rm T}u^1,\\
u^0{u^1}^{\rm T}&=&q^2{u^1}^{\rm T}u^0,\\
u^1{u^1}^{\rm T}&=&q{u^1}^{\rm T}u^1.                       \label{poincare2}
\end{eqnarray}
The quantum determinant $\sigma$ is still defined by Eq.~(\ref{det}), but now
$\sigma$ is not a central element of $P_q(2)$.

\subsection{Hopf Structure}
For the completness we show also the Hopf structure of our $P_q(2)$.

The co-product is the following
\begin{equation}
\Delta\pmatrix{a&b&u^0\cr q^{-1}b^{\rm T}&a^{\rm T}&u^1\cr0&0&{\bf1}}
=\pmatrix{a&b&u^0\cr q^{-1}b^{\rm T}&a^{\rm T}&u^1\cr0&0&{\bf1}}
\otimes\pmatrix{a&b&u^0\cr q^{-1}b^{\rm T}&a^{\rm T}&u^1\cr0&0&{\bf1}},
                                                           \label{co-product}
\end{equation}
where $\otimes$ means tensor product and usual matrix multiplication together.

The co-unity:
\begin{equation}
\epsilon\pmatrix{a&b&u^0\cr q^{-1}b^{\rm T}&a^{\rm T}&u^1\cr0&0&{\bf1}}
=\pmatrix{1&0&0\cr0&1&0\cr0&0&1};
\end{equation}
and finally the antipode:
\begin{eqnarray}
\kappa\pmatrix{a&b&u^0\cr q^{-1}b^{\rm T}&a^{\rm T}&u^1\cr0&0&{\bf1}}
&=&\sigma^{-1}\pmatrix{a^{\rm T}&-q^{-1}b&0\cr -b^{\rm T}&a&0\cr0&0&{\bf1}}
\pmatrix{{\bf1}&0&-u^0\cr0&{\bf1}&-u^1\cr0&0&\sigma}
                                                                    \nonumber\\
&=&
\pmatrix{\sigma^{-1}a^{\rm T}&-q^{-1}\sigma^{-1}b&-\sigma^{-1}a^{\rm T}u^0\cr
-\sigma^{-1}b^{\rm T}&\sigma^{-1}a&-\sigma^{-1}au^1\cr
0&0&{\bf1}}.                                                 \label{antipode}
\end{eqnarray}
The structure for $L_q(2)$ is the same as above, with cut off the last rows and
columns and putting $\sigma={\bf1}$.

\section{Quantum Minkowski Space-Time}
Now we return to the quantum space-time described in sect.~2.

\subsection{Algebraic Structure of Quantum Space-Time}
Till now we have not fixed commutatiuons relations between $x^\mu$'s and
${x^\mu}^{\rm T}$'s, except the relation Eq.~(\ref{reorder}). But the
requiremet of covariance of the algebra ${\bf E}^{1,1}_q$ under the action of
$P_q(2)$ induces the following rules.
\begin{eqnarray}
x^0x^1&=&qx^1x^0,                                               \label{mink1}\\
x^0{x^0}^{\rm T}&=&q{x^0}^{\rm T}x^0+q(q^2-1){x^1}^{\rm T}x^1,\\
x^0{x^1}^{\rm T}&=&q^2{x^1}^{\rm T}x^0,\\
x^1{x^1}^{\rm T}&=&q{x^1}^{\rm T}x^1.                           \label{mink2}
\end{eqnarray}
Thus, the Lorentz invariant transforms uder the action of the Poincar\'e group
in the following way
\begin{equation}
\delta({\mit\Sigma})=\sigma\otimes{\mit\Sigma},
\end{equation}
so we can see the rescaling of the invariant.

We can attract this rescaling by a redefinition of the Lorentz invariant.
Instead of Eq.~(\ref{metrics}) we choose
\begin{eqnarray}
{\mit\Sigma}^2&=&{x^0}^{\rm T}{\mit\Gamma} x^0-{x^1}^{\rm T}{\mit\Gamma} x^1
                                                                    \nonumber\\
&=&\pmatrix{{x^0}^{\rm T}&{x^1}^{\rm T}}
\pmatrix{{\mit\Gamma}&0\cr0&-{\mit\Gamma}}
\pmatrix{x^0\cr x^1},                                       \label{g-metrics}
\end{eqnarray}
where ${\mit\Gamma}$ is an additional generator in ${\bf E}^{1,1}_q$,
${\mit\Gamma}^{\rm T}={\mit\Gamma}$, $\bar{{\mit\Gamma}}={\mit\Gamma}$, and
obeys the following commutation rules
\begin{equation}
{\mit\Gamma} x^\mu=q^3x^\mu{\mit\Gamma}.
\end{equation}
If we assume that
\begin{equation}
\delta({\mit\Gamma})=\sigma^{-1}\otimes{\mit\Gamma},
\end{equation}
we get no rescaling of ${\mit\Sigma}^2$.

Independently of choice of definition of the metrics (Eq.~(\ref{metrics}) or
Eq.~(\ref{g-metrics})) we can find for squares of generators defined by
Eq.~(\ref{hermicity})
\begin{equation}
{x^0}^2{x^1}^2={x^1}^2{x^0}^2,                                     \label{xx}
\end{equation}
so, however coordinates do not commute, observables commute, then we can
measure them together at the same moment.

\subsection{Differential Calculus on Quantum Space-Time}
Our next point is to find a differential calculus on ${\bf E}^{1,1}_q$. In our
investigations we will follow with definition of differential calculus given in
\cite{holomorph}. Let ${\rm d}$ will be a map ${\rm d}:{\bf E}^{1,1}_q\to {\rm
d}{\bf E}^{1,1}_q$, where ${\rm d}{\bf E}^{1,1}_q$ is an algebra ${\bf
E}^{1,1}_q$ extended by the sector generated by ${\rm d}x^\mu$'s and their
transposition, ${\rm d}$ obeys the following rules:
\begin{enumerate}
\item linearity; \item graded Leibniz rule; \item nilpotency.
\end{enumerate}
Next, we assume an existence of an affine parameter $\tau$, such that we can
write $x^\mu=x^\mu(\tau)$ and ${\rm d}x^\mu=\dot x^\mu(\tau){\rm d}\tau$,
i.e.\ $\dot x^\mu$ is standard (not quantum) derivative $x^\mu$ on $\tau$.
The affine parameter $\tau$ in a classical limit has an interpretation as the
proper time.

Under the assumptions mentioned above we can find the following consistent,
associative differential calculus on ${\bf E}^{1,1}_q$ (in the language of the
genertors and their derivatives)
\begin{eqnarray}
x^0\dot x^0&=&q^{-2}\dot x^0x^0,                                \label{diff1}\\
x^0\dot x^1&=&q^{-1}\dot x^1x^0,\\
x^0{\dot x^0{}}^{\rm T}&=&q{\dot x^0{}}^{\rm T}x^0
                         +q(q^2+1){\dot x^1{}}^{\rm T}x^1,\\
x^0{\dot x^1{}}^{\rm T}&=&{\dot x^1{}}^{\rm T}x^0,
\end{eqnarray}
\begin{eqnarray}
x^1\dot x^0&=&q^{-1}\dot x^0x^1-(1-q^{-2})\dot x^1x^0,\\
x^1\dot x^1&=&q^{-2}\dot x^1x^1,\\
x^1{\dot x^0{}}^{\rm T}&=&q^2{\dot x^0{}}^{\rm T}x^1,\\
x^1{\dot x^1{}}^{\rm T}&=&q{\dot x^1{}}^{\rm T}x^1,
\end{eqnarray}
\begin{eqnarray}
\dot x^0\dot x^1&=&q\dot x^1\dot x^0,\\
\dot x^0{\dot x^0{}}^{\rm T}&=&q{\dot x^0{}}^{\rm T}\dot x^0
                              +q(q^2-1){\dot x^1{}}^{\rm T}\dot x^1,\\
\dot x^0{\dot x^1{}}^{\rm T}&=&q^2{\dot x^1{}}^{\rm T}\dot x^1,\\
\dot x^1{\dot x^1{}}^{\rm T}&=&{\dot x^1{}}^{\rm T}\dot x^1.
                                                                \label{diff2}
\end{eqnarray}
It is easy to see that the above differential calculus is covariant under the
action of the quantum Lorentz group $L_q(2)$.

\section{Free Motion}
We want to find a dependence of the generators of ${\bf E}^{1,1}_q$ on the
affine parameter $\tau$ for the class of inertial frames of the reference. As
the inertial frame we will understand the frame wich we can obtain from the
rest system (i.e.\ system with $\dot x^1=0$) by a Poincar\'e transformation.

\subsection{The Rest Solution}
It is easy to check that the condition of the linear dependence $x^0$ on $\tau$
(which is the most natural in the classical case) contradicts with our
differential calculus. Indeed, from the Eq.~(\ref{diff1}) it is evident that
the Leibniz rule gives us
\begin{equation}
x^0\ddot x^0=q^{-2}\ddot x^0x^0+(q^{-2}-1)(\dot x^0)^2,
\end{equation}
(here $(\dot x^0)^2=\dot x^0\dot x^0$, not ${\dot x^0{}}^{\rm T}\dot x^0$) and
for the linear dependence we have $\ddot x^0=0$, what implies $\dot x^0=0$ and
no dependence $x^0$ on $\tau$ at all!

In the rest frame we have the following algebra of the generator of ${\bf
E}^{1,1}_q$
\begin{eqnarray}
x^0_R{x^0_R}^{\rm T}&=&q{x^0_R}^{\rm T}x^0_R,                   \label{rest1}\\
x^0_R\dot x^0_R&=&q^{-2}\dot x^0_Rx^0_R,                        \label{rest2}\\
x^0_R{\dot x^0_R{}}^{\rm T}&=&q{\dot x^0_R{}}^{\rm T}x^0_R,     \label{rest3}\\
\dot x^0_R{\dot x^0_R{}}^{\rm T}&=&q{\dot x^0_R{}}^{\rm T}\dot x^0_R,
                                                                \label{rest4}
\end{eqnarray}
here we put $x^1_R=0$ for simplicity, the case with non-zero $x^1_R$ we can
get from this case by a translation in the direction $x^1$.
We assume an unitarity of the developement of $x^0_R$ on $\tau$.
\begin{eqnarray}
x^0_R(\tau)&=&U(\tau)x^0_R(0)U^\dagger(\tau),                   \label{unit1}\\
\dot x^0_R(\tau)&=&U(\tau)\dot x^0_R(0)U^\dagger(\tau),         \label{unit2}
\end{eqnarray}
where $UU^\dagger=u^\dagger U=I$ and $x^0_R(0)$, $\dot x^0_R(0)$ are constant
generators. We will denote them by the other pair of constant generators as
\begin{eqnarray}
x^0_R&=&X^0_R,                                                 \label{const1}\\
\dot x^0_R&=&{\mit\Omega}X^0_R,                                \label{const2}
\end{eqnarray}
and the algebra Eq.~(\ref{rest1}--\ref{rest4}) is of the following form
\begin{eqnarray}
X^0_R{X^0_R}^{\rm T}&=&{X^0_R}^{\rm T}X^0_R,                    \label{Rest1}\\
x^0_R{\mit\Omega}&=&q^{-2}{\mit\Omega}X^0_R,\\
x^0_R{\mit\Omega}^{\rm T}&=&{\mit\Omega}^{\rm T}X^0_R,\\
{\mit\Omega}{\mit\Omega}^{\rm T}&=&{\mit\Omega}^{\rm T}{\mit\Omega}.
                                                                \label{Rest2}
\end{eqnarray}
We can also rewrite the unitarity condition Eq.~(\ref{unit1}--\ref{unit2}) in
the form
\begin{eqnarray}
x^0_R(\tau)&=&F({\mit\Omega},\tau)X^0_R,                       \label{Funit1}\\
{x^0_R}^{\rm T}(\tau)&=&{X^0_R}^{\rm T}F({\mit\Omega}^{\rm T},\tau),
                                                               \label{Funit2}
\end{eqnarray}
where the Function $F$ fulfils the boundary condition $F({\mit\Omega},0)=1$.

It is easy to see that in the above parametrization the
Eq.~(\ref{rest1}, \ref{rest3}, \ref{rest4}) holds automatically, but
Eq.~(\ref{rest2}) gives us
\begin{equation}
\dot F({\mit\Omega},\tau)F(q^{-2}{\mit\Omega},\tau)
=q^2F({\mit\Omega},\tau)\dot F(q^{-2}{\mit\Omega},\tau),        \label{equat}
\end{equation}
i.e.\ differential equation on $F$ as a function of $\tau$.

The Eq.~(\ref{equat}) has a general solution
\begin{equation}
F({\mit\Omega},\tau)=\exp[{\mit\Omega}\ln(\lambda(\tau))],      \label{solut}
\end{equation}
where $\lambda(\tau)$ is any number-valued function of $\tau$, with condition
$\lambda(0)=1$.

Finally we come into the parametrization of $x^0_R$
\begin{eqnarray}
x^0_R(\tau)&=&\exp[{\mit\Omega}\ln(\lambda(\tau))]X^0_R,       \label{restx0}\\
\dot x^0_R(\tau)&=&\frac{\dot\lambda(\tau)}{\lambda(\tau)}
                  {\mit\Omega}\exp[{\mit\Omega}\ln(\lambda(\tau))]X^0_R.

                                                      \label{restx.0}
\end{eqnarray}
It is evident that in the classical case (${\mit\Omega}\to1$) we obtain
$F(\tau)=\lambda(\tau)$, i.e.\ we can parametrize $x^0_R$ by any function
of $\tau$.

\subsection{Free Motion Solution}
The general solution for the free motion we can obtain in the form
\begin{eqnarray}
x^0(\tau)&=&\exp[{\mit\Omega}\ln(\lambda(\tau))]X^0,          \label{freex0}\\
x^1(\tau)&=&\exp[{\mit\Omega}\ln(\lambda(\tau))]X^1,          \label{freex1}
\end{eqnarray}
where
\begin{eqnarray}
X^0&=&a\otimes X^0_R,                                              \label{X0}\\
X^1&=&q^{-1}b^{\rm T}\otimes X^0_R,                                \label{X1}
\end{eqnarray}
or if we admit translations
\begin{eqnarray}
X^0&=&a\otimes X^0_R+u^0\otimes{\bf 1},                           \label{X0t}\\
X^1&=&q^{-1}b^{\rm T}\otimes X^0_R+u^1\otimes{\bf 1}.             \label{X1t}
\end{eqnarray}
In both case, of course, $X^0$ and $X^1$ are constant in $\tau$.

We would note that the above definition of free motion means that the
two-velocity (the $D=2$ analogon of four-velocity) is given by
\begin{equation}
\dot X^\mu=\frac{\dot\lambda(\tau)}{\lambda(\tau)}{\mit\Omega}
           \exp[{\mit\Omega}\ln(\lambda(\tau))]X^\mu,
\end{equation}
while the one-velocity (the $D=2$ analogon of three-velocity) we should define
as the quantum (Gauss-Jackson) derivative of $x^1$ on $x^0$.

\section{Momenta and Phase-Space}
To finish our discussion about kinematics we want to define the momenta and
mass.

We follow with the standard definition of the two-momentum
\begin{eqnarray}
p^0&=&m\dot x^0,                                                   \label{p0}\\
p^1&=&m\dot x^1,                                                   \label{p1}
\end{eqnarray}
i.e.\ the two-momenta are proportional to two-velocity and the proportionality
coefficient we will call the mass.

\subsection{Discussion About the Mass}
We assume that the mass is the generator which transforms in the following way
\begin{equation}
\delta(m)=\sigma\otimes m.
\end{equation}
i.e.\ like a scalar under the action of the $q$-Poincar\'e group.

The conservation of the momenta $\dot p^\mu=0$ leads to the dependence $m$ on
$\tau$
\begin{equation}
m(\tau)=\frac{\lambda(\tau)}{\dot\lambda(\tau)}M
        \exp[-{\mit\Omega}\ln(\lambda(\tau))],                   \label{mass}
\end{equation}
where $M$ is a constant generator.

Assuming the {\em Bethe Ansatz\/} between $m$ and $\dot x^\mu$
\begin{equation}
m\dot x^\mu=\xi\dot x^\mu m,                                       \label{xi}
\end{equation}
we get
\begin{eqnarray}
M{\mit\Omega}&=&q^2{\mit\Omega}M,                                  \label{M1}\\
MX^\mu&=&\xi q^{-2}X^\mu M.                                        \label{M2}
\end{eqnarray}
For the other constant generators we assume the {\em Bethe Ansatz\/} too:
\begin{eqnarray}
M^{\rm T}{\mit\Omega}&=&\omega{\mit\Omega}M^{\rm T},              \label{Mt1}\\
M^{\rm T}X^\mu&=&\theta X^\mu M^{\rm T},                          \label{Mt2}
\end{eqnarray}
the consistency requires
\begin{equation}
MM^{\rm T}=(\omega\theta)^2M^{\rm T}M.                             \label{MM}
\end{equation}
Eqs.~(\ref{xi}--\ref{MM}) lead to the following algebra
\begin{eqnarray}
m\dot x^\mu&=&\xi\dot x^\mu m,                                    \label{mx.}\\
m{\dot x^\mu{}}^{\rm T}&=&\theta^{-1}{\dot x^\mu{}}^{\rm T}m
                          \exp[(1-\omega){\mit\Omega}^{\rm T}\ln(\lambda)],
                                                                 \label{mx.t}
\end{eqnarray}
\begin{eqnarray}
mx^\mu&=&\xi q^{-2}x^\mu m,                                        \label{mx}\\
m{x^\mu}^{\rm T}&=&\omega\theta^{-1}{x^\mu}^{\rm T}m
                          \exp[(1-\omega){\mit\Omega}^{\rm T}\ln(\lambda)].
                                                                  \label{mxt}
\end{eqnarray}
Note that the Eqs.~(\ref{mx.t}) and (\ref{mxt}) are in the {\em Bethe Ansatz\/}
only if we choose $\omega=1$. Hwnceforth we will deal with this choice.

\subsection{The Algebra of Phase-Space}
The results of previous subsection allow us to write the set of all of the
commutation relations in the phase-space.  Together with the
Eqs.~(\ref{mink1}--\ref{mink2}) we have
\begin{eqnarray}
x^0p^0&=&\xi^{-1}p^0x^0,                                       \label{phase1}\\
x^0p^1&=&\xi^{-1}qp^1x^0,\\
x^0{p^0}^{\rm T}&=&\theta^{-1}{p^0}^{\rm T}x^0
                   +\theta^{-1}q(q^2-1){p^1}^{\rm T}x^1,\\
x^0{p^1}^{\rm T}&=&\theta^{-1}q^2{p^1}^{\rm T}x^0,
\end{eqnarray}
\begin{eqnarray}
x^1p^0&=&\xi^{-1}qp^0x^1
         -\xi^{-1}(q^2-1)p^1x^0,\\
x^1p^1&=&\xi^{-1}p^1x^1,\\
x^1{p^0}^{\rm T}&=&\theta^{-1}q^2{p^0}^{\rm T}x^1,\\
x^1{p^1}^{\rm T}&=&\theta^{-1}q{p^1}^{\rm T}x^1,
\end{eqnarray}
\begin{eqnarray}
p^0p^1&=&qp^1p^0,\\
p^0{p^0}^{\rm T}&=&q{p^0}^{\rm T}p^0
                   +q(q^2-1){p^1}^{\rm T}p^1,\\
p^0{p^1}^{\rm T}&=&q^2{p^1}^{\rm T}p^0,\\
p^1{p^1}^{\rm T}&=&q{p^1}^{\rm T}p^1.                          \label{pahse2}
\end{eqnarray}
Notice that the algebra of the components of momenta is exactly tha same as the
algebra of the components of coordinates, as it is expectable, since
coordinates and momenta are the two-vectors under the action of the quantum
Poincar\'e group.

\subsection{The ``Uncertainity Relations''}
We can construct the hermitian elements from $p^\mu$'s via the same way as for
$x^\mu$'s
\begin{equation}
{p^\mu}^2={p^\mu}^{\rm T}p^\mu
\end{equation}
and interprete them as the squares of the energy (${p^0}^2$) and of
the linear momentum (${p^1}^2$).

We have also
\begin{equation}
{p^0}^2{p^1}^2={p^1}^2{p^0}^2,
\end{equation}
like for $x$'s.

The most interesting are the relations between $x$'s and $p$'s. They are the
following
\begin{eqnarray}
{x^0}^2{p^1}^2&=&{p^1}^2{x^0}^2,                                 \label{x0p1}\\
{x^1}^2{p^1}^2&=&{p^1}^2{x^1}^2,                                 \label{x1p1}\\
{x^0}^2{p^0}^2&=&{p^0}^2{x^0}^2
                 +(q^2-1)

\left({x^0}^{\r
m T}x^1{p^1}^{\rm T}p^0

-{p^0}^{\
rm T}p^1{x^1}^{\rm T}x^0\right),

\labe
l{x0p0}\\
{x^1}^2{p^0}^2&=&{p^0}^2{x^1}^2
                 +\xi\theta^{-1}(q^2-1)

\left({p^1}^{\r
m T}{x^0}^{\rm T}x^1p^0

-{p^0}^{\
rm T}{x^1}^{\rm T}x^0p^1\right).

                                                   \label{x1p0}
\end{eqnarray}
So we have the ``uncertainity relation'' between the energy and time and
between the energy and position, what is completely unexpectable.  For the
other hand we {\em have not\/} the uncertainity relation between the momentum
and position.

Notice that this non-commutativity is in the second order $q$ and for $q$ from
the neighbourhood of $1$ tends very fast to $0$.

\section*{Acknowledgements}
I am grateful to J.~Rembieli\'nski for many interesting and helpful discussions
and comments. I also wolud like to thank to the Organizers of the ``Conference
on Quantum Topology'', especially to D.~Yetter, for kind hospitality.

\end{document}